\documentstyle[11pt,twoside,psfig,jltp]{article}

\title{Quantum nature of dislocations in pure bcc Helium}

\author{Nir Gov\address{Department of Physics, 
University of Illinois at Urbana-Champaign, \\
1110 Green St., Urbana 61801, IL, USA}}

\runninghead{N. Gov}{Quantum nature of dislocations in pure bcc Helium}
       
\begin{document}

\begin{abstract}
Recent experiments show the thermal growth of dislocation lines in unlta-pure bcc $^{3}$He. The activation energy for the growth of the dislocation lines is found to agree with the activation energy of mass diffusion. We propose that these dislocations are topological defects in the phase of the complex order-parameter which describes the dynamic zero-point atomic correlations, unique to the bcc phase.
These is a shear strain field associated with these topological defects.
We show that the smallest topological defect is a localized excitation, a loop-defect, which leads to the exponential growth of the dislocation lines with temperature.

\end{abstract}

\maketitle

\vspace{0.3in}

Recent experiments \cite{miura} show the thermal growth of dislocation lines in ultra-pure bcc $^{3}$He, interpreted within the dislocation theory of Granato and L\"ucke \cite{granato}. The activation energy for the growth of dislocation network is found to agree with the activation energy of mass diffusion \cite{e0he3}. 
This is in contrast with the case of hcp $^4$He, where only pinning of dislocations to $^3$He impurities was observed \cite{paalanen}.
It is the quantum nature of the dislocations in the pure bcc phase which we shall describe in this paper. 

We have previously proposed a new model for the description of the quantum correlations in the bcc phase of solid Helium \cite{niremil,nirbcc}. This model introduces a new complex order parameter which describes the phase of the dynamic zero-point oscillations of the atoms. These zero-point oscillations are highly directional along the normal ((100),(010),(001)) axes of the bcc phase and result in dynamic polarization of the electronic clouds, i.e. $s-p$ level mixing of the order of $\sim1\%$.
The oscillating electric dipole moments are correlated in a long-range order to lower the ground-state energy (Fig.1).
This is a state with Off-Diagonal Long Range Order (ODLRO) \cite{nirbcc}, with a broken gauge symmetry in the form of a global phase of the oscillating dipoles.
A flipping of a dipole out of this ordered ground-state, i.e. giving it a phase shift of $\pi$, is treated as a local-mode of energy $E_{0}$, which is twice the dipolar interaction energy.
The dipoles are simultaneously arranged along the three orthogonal axes, and form a coherent state with a complex order-parameter having 3 independent phases for each of the normal axes \cite{nirbcc}.
In the hcp phase we do not expect such a long-range coherent order to form since the triangular symmetry frustrates the dipolar interactions

\begin{figure}
\centerline{\psfig{file=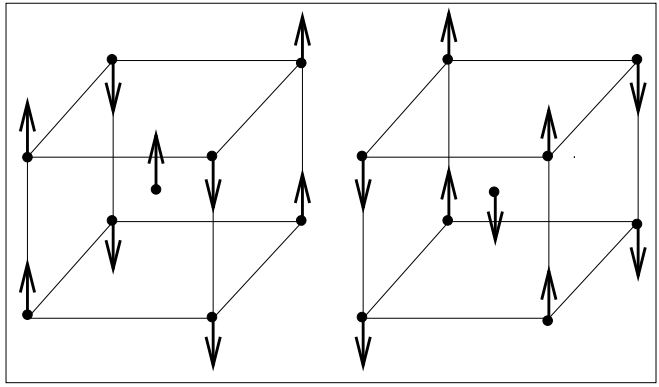,height=2.0in}}
\caption{The coherent long-range order of the zero-point dipoles in the bcc phase. The instantaneous zero-point dipoles are shown along one of the three normal axes, oscillating between these two configurations (quantum resonance) with frequency $\hbar/E_{0}$.}
\end{figure}

The effect of these interacting local dipoles is to couple (hybridize) with the harmonic transverse phonon which has the same symmetry as the dipolar order of Fig.1, which turns out to be only the T$_{1}$(110) phonon. The result of this hybridization \cite{niremil} is to soften this harmonic phonon by a factor of 2, and create a new localized (i.e. dispersionless) excitation of twice the bare local-mode energy: $2E_{0}$.
The nature of this excitation \cite{niremil} is shown in Fig.2. This excitation involves a local flipping of the phase of the dipoles, which is spread over approximately two unit cells. The resulting "breathing" motion will lower the barrier for atoms to exchange and this excitation is therefore involved in the thermal activation of mass diffusion in bcc Helium.
Since the T$_{1}$(110) phonon is a shear wave, the flipping of a dipole out of the ground-state, introduces a local shear.
The shear degree of freedom is introduced into the hybridization through the harmonic transverse phonon \cite{niremil}.

We will now show that the localized excitation of energy $2E_{0}$ is a shear-dipole, i.e. a microscopic analogue of a vortex-loop.
In the hybridization procedure of the dipoles and the harmonic T$_{1}$(110) shear phonon \cite{niremil}, we found that the two excitation branches have ground-state occupation of local-modes (dipole flips) and harmonic phonons
\begin{eqnarray}
\left| 0_1\right\rangle  &=&\prod_k\exp \left( \frac {C(k)}{A(k)}{a^{\dagger }}_k{%
a^{\dagger }}_{-k}\right) \exp \left( \frac {D(k)}{B(k)}b{^{\dagger }}_kb{^{\dagger }}%
_{-k}\right)   \nonumber \\
\left| 0_2\right\rangle  &=&\prod_k\exp \left( \frac {D(k)}{B(k)}{a^{\dagger }}_k{%
a^{\dagger }}_{-k}\right) \exp \left( \frac {C(k)}{A(k)}b{^{\dagger }}_kb{^{\dagger }}%
_{-k}\right)   \label{psi0hyb}
\end{eqnarray}
where the functions $A(k),B(k),C(k),D(k)$ are described in [\cite{niremil,hopfield}], and the operators $a^{\dagger }_k,a_k$ and $b{^{\dagger }}_k,b_k$ are the creation/annihilation operators of the local-mode and harmonic phonon respectively. The subscript 1 and 2 stands for the soft (i.e. hybridized) T$_{1}$(110) phonon and the localized $2E_{0}$ branch respectively.
From (\ref{psi0hyb}) the occupation number of local-modes in the lower branch diverges at $k\rightarrow 0$ as: $\left\langle n_{a}(k)\right\rangle=|C(k)/A(k)|^2\rightarrow(3E_{0}/8)/\hbar k c$, where $c$ is the velocity of the soft T$_{1}$(110) phonon.
This signals condensation in a Bose system \cite{nirbcc}, which in this low energy limit shows the ODLRO in this system.
Similarly the number of harmonic shear-modes occupation diverges for $k\rightarrow 0$ in the localized branch.
Such occupation of pairs of $(k,-k)$ shear modes (\ref{psi0hyb}) with $1/k$ divergency is equivalent to a $1/r^2$ shear field, which coincides with the large $r$ field around a localized shear dipole (Fig.2).

\begin{figure}[tbp]
\input epsf \centerline{\ \epsfysize 3.0cm \epsfbox{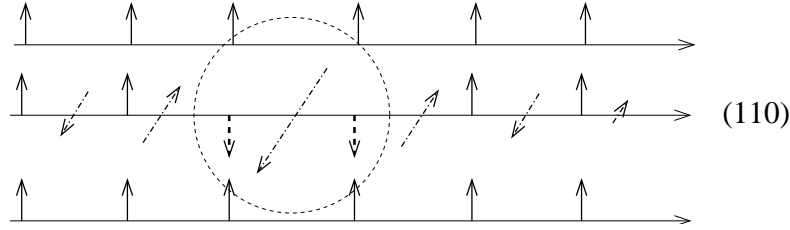}}
\caption{The dipolar nature of the localized excitation
$2E_{0}$
, as two adjacent local modes (i.e. dipole-flips, circled dashed arrows) along the (110) direction. The shear-stress field
is indicated by the dash-dot slanted arrows.}
\end{figure}

We shall now turn to describe topological defects of macroscopic length in the dipolar order of Fig.1.
In general, a complex order-parameter can support linear topological defects around which the coherent phase changes by $2n\pi$.
 This is the case in superfluid $^4$He and s-wave superconductors \cite{tilley}. In our case we can treat each component of the 3 orthogonal dipoles as an independent order-parameter. 
Concentrating on one axis, a linear topological defect appears as a complex phase in the dipole-dipole interaction energy of the atom at ${\bf r}_{0}$
\begin{equation}
E_{dipole}=-\left| {\bf \mu }\right| ^{2}\sum_{i\neq 0}\left[ \frac{3\cos
^{2}\left( {\bf \mu }\cdot \left( {\bf r}_{0}-{\bf r}_{i}\right) \right) -1}{%
\left| {\bf r}_{0}-{\bf r}_{i}\right| ^{3}}\right]  
\exp \left[ i\cdot \Theta\left( {\bf r}_{vort}-{\bf r}_{i}\right) \right]
\label{edipole}
\end{equation} 

where the sum is over all the atoms in the lattice, ${\bf r}_{i}$ being the
coordinate of the $i$-th atom and ${\bf \mu }$ is the coherently oscillating electric dipole moment along one of the normal axes \cite{niremil,nirbcc}. The geometrical phase $\Theta\left( {\bf r}_{vort}-{\bf r}_{i}\right)$ is just the azimuthal angle and changes by $2n\pi$ around the defect's axis (located at ${\bf r}_{vort}$).
We have calculated by numerical summation (\ref{edipole}) the radial dependence of the dipolar energy cost per atom, for various linear defects in the bcc lattice (Fig.3), with respect to the ground-state of Fig.1. 
We found for a defect with $n=1$, that the energy cost per atom at radius $R$ is
\begin{equation}
E_{defect}(R)= \alpha |E_{dipole}| (a/R)^2/2 
\label{edipolecont}
\end{equation}
The values of the numerical prefactor $\alpha$ will be different for various defects, due to the dipolar energy
form (\ref{edipole}), and are given in Fig.3.
The form of the energy (\ref{edipolecont}) is what we expect from a continuum model where the dipolar energy is simply proportional to the scalar product of the nearest-neighbor dipoles.
This form is the same as that for the kinetic energy around a linear vortex in superfluid $^4$He \cite{donnely}.

\begin{figure}
\centerline{\psfig{file=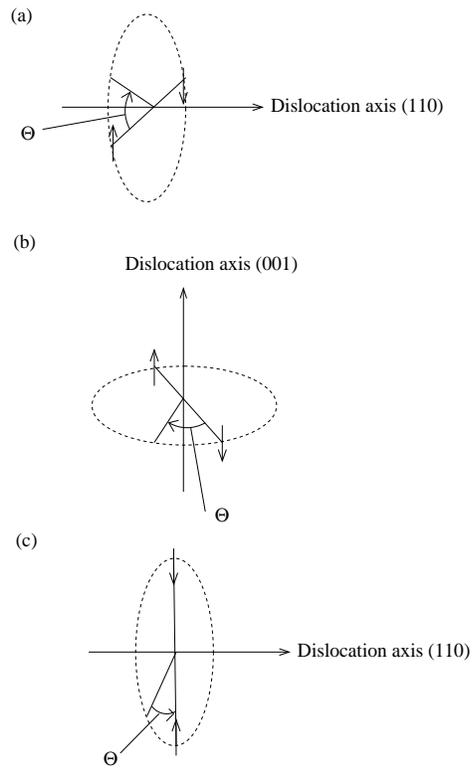,height=4.0in}}
\caption{The linear topological defects in the dipolar array of the bcc phase. The dipoles show the phase changing by $2\pi$ around the defect's axis, with the azimuthal angle $\Theta$. The numerical prefactor $\alpha$ (in Eq.(\ref{edipolecont})) is: (a) 1 (b) 1/4 (c) 3/4.}
\end{figure}

In addition to linear defects, which have to terminate at the system's surface, there can be loop-defects (again similar to vortex-loops in superfluid \cite{donnely} $^4$He).
The energy cost of a loop-defect of radius $R$ in a continuum model \cite{donnely} is
\begin{equation}
E_{loop}=\pi^2 E_{0} \left(2R/a \right) ln(R/a)
\label{eloop}
\end{equation}

where the local-mode energy is defined to be \cite{niremil,nirbcc}: $E_{0}=-2E_{dipole}$.
The smallest such loop should have radius of approximately one unit cell, where the continuum limit is already suspect. Fortunately in our previous treatment we have found a localized excitation of exactly the nature of the smallest loop-defect.
This is the excitation we have discussed above (Fig.2), in relation to mass-diffusion. The dipolar phase changes by $2\pi$ (with respect to the ground-state phase) when passing through this excitation along the (110) direction. The dipolar phase changes by $\pi$ due to the first dipole-flip and then by a second $\pi$ due to the second dipole-flip (Fig.4). It has the form of a topological defect of the kind marked (a) in Fig.3, curled into the smallest possible loop.
In addition, using this excitation's energy $2E_{0}$ in (\ref{eloop}) we get a loop radius of $R\sim 1.1a$, as we expect from the smallest loop-defect.

We now come to the recent experimental results \cite{miura}. These results show that the dislocation network and loops have a length which grows exponentially with temperature, even in ultrapure crystals.
The activation energy for this dislocation growth is $\sim3$K at V=24.3 cm$^3$/mole. This energy agrees with the value $2E_{0}$, as measured in various experiments for mass diffusion and specific-heat \cite{e0he3}.
We therefore conclude that the dislocations measured in ultrapure bcc $^3$He are in fact topological defects in the phase of the coherent zero-point dipoles.
These linear defects will grow and lengthen due to thermal activation of small loop-defects of energy $2E_{0}$ (Fig.4).

\begin{figure}
\centerline{\psfig{file=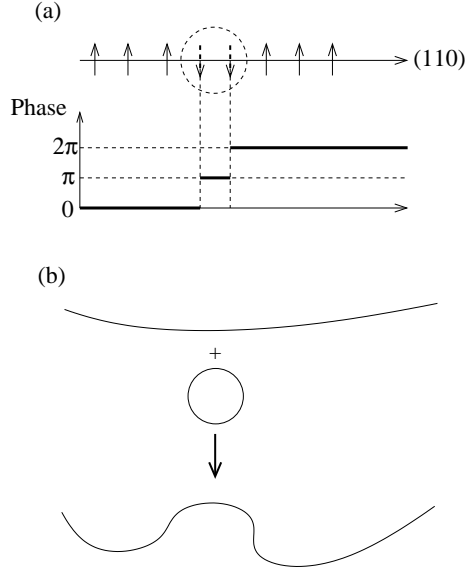,height=3.0in}}
\caption{(a) Loop defects along the (110) direction as equivalent to the localized excitation of energy $2E_{0}$. The phase changes twice by $\pi$ across this defect. (b) Thermally activated loop-defects will increase the overall length of the network of defects, through processes such as reconnection.}
\end{figure}

In the ultrapure hcp phase there was no measurement of such effect, only the weak influence of thermal phonons \cite{paalanen}. This difference between the two phases fits our interpretation, since due to geometric frustration we do not expect a similar complex order-parameter to exist in the hcp phase \cite{niremil}.

To summarize, we have shown that the novel phase of bcc He, which has a complex order-parameter, can support topological defects of microscopic and macroscopic sizes.
These defects will behave similar to a network of dislocation lines, and their total length increase with the thermal activation of microscopic loop-defects.
This explains the different experimental measurements in bcc and hcp phases.
To conclude we point to possible experimental verifications of the proposed model. 
It follows from our description that in ultrapure bcc $^3$He at different molar volumes the dislocation network is thermally activated with the same energy as found in mass-diffusion and specific-heat measurements \cite{e0he3}. This prediction can be checked in future experiments.
Finally, in the spin-ordered bcc $^3$He, we expect the topological defects to have a nuclear spin-structure, due to the proposed relation between the zero-point atomic motion and the nuclear spin-ordering phenomenon \cite{bcche3}. These dislocations may be therefore detectable using magnetic probes such as NMR.

\section*{ACKNOWLEDGMENTS}
This work was supported by 
the Fulbright Foreign Scholarship grant, 
NSF grant no. DMR-99-86199  and
NSF grant no. PHY-98-00978.

\end{document}